\documentclass[a4paper,11pt]{article}
\usepackage[margin=2.35cm]{geometry}
\usepackage{amsmath,amssymb,bm}
\usepackage{booktabs}
\usepackage{graphicx}
\usepackage{xcolor}
\usepackage[authoryear,round]{natbib}
\usepackage[hidelinks]{hyperref}
\providecommand{\doi}[1]{doi:\space\href{https://doi.org/#1}{\nolinkurl{#1}}}
\numberwithin{equation}{section}
\title{\textcolor{black}{Long-term average collision frequencies between two bodies undergoing Lidov--Kozai evolution}}
\author{%
\parbox{0.97\textwidth}{\centering
Youpeng Liang\textsuperscript{1} and
Xiaodong Liu\textsuperscript{1,2}\thanks{Corresponding author. E-mail: \href{mailto:liuxd36@mail.sysu.edu.cn}{liuxd36@mail.sysu.edu.cn}}\\[0.7em]
{\small
\textsuperscript{1}School of Aeronautics and Astronautics, Shenzhen Campus of Sun Yat-sen University, Shenzhen, Guangdong 518107, China\\[0.35em]
\textsuperscript{2}Shenzhen Key Laboratory of Intelligent Microsatellite Constellation, Shenzhen, Guangdong 518107, China
}}}
\date{}

\begin{document}

\maketitle

\begin{abstract}
\textcolor{black}{The existing \"Opik-type collision frequency framework that accounts for the projectile's Lidov--Kozai (LK) secular evolution holds the target's semimajor axis, eccentricity, and inclination fixed and prescribes uniform precession of its longitude of the ascending node and argument of periapsis.
This treatment is not readily applicable when the target's eccentricity, inclination, or orbital precession rates undergo appreciable secular variations. Here, we extend the framework of \citet{liang4} to address one such case, in which both the projectile and target undergo LK evolution.
Using a doubly averaged quadrupole model to describe the secular dynamics of both bodies, we include the target's orbital evolution when evaluating the collision contribution of each orbital intersection configuration.
In the numerical example considered, the projectile survival curve predicted using the calculated collision frequency shows an overall decay trend close to that obtained from direct N-body simulations using REBOUND. In addition, we examine whether the long-term mean collision frequency depends on initial states while holding both bodies’ semimajor axes, Kozai constants and closed Hamiltonian level curves fixed. Our theoretical analysis indicates that this frequency is independent of initial states when the parameters describing their relative state are jointly equidistributed at the start of successive projectile LK cycles. Numerical cases with different initial states support this conclusion.}
\end{abstract}

\section{Introduction}\label{p5s1:first}

Collisions are an important process in planetary science: they can influence how populations of celestial bodies evolve, alter their size distributions, and facilitate material transfer between bodies.
Estimates of collision frequencies have been used to study the collisional evolution of the asteroid belt \citep{cib2014}, estimate the meteoroid impact flux at Earth \citep{nes2010}, and analyze the deposition of dust released by irregular satellites on the surfaces of other satellites \citep{tam2011}.
Perturbations drive secular changes in the shapes and orientations of both orbits, causing their relative orbital configuration and associated collision probability to evolve over time. 
Calculating the long-term average collision frequency therefore requires accounting for the collision contributions from the orbital intersection configurations encountered during the secular evolution of both bodies.

Vokrouhlický et al. (2012) developed a semi-analytical Öpik-type method for calculating the collision probabilities of projectiles undergoing Lidov–Kozai (LK) secular evolution.
Unlike traditional \"Opik-type treatments that hold the projectile's eccentricity and inclination fixed, this method allows both quantities to undergo the secular variations associated with LK evolution.
The target, however, is restricted to a circular orbit in the perturber's orbital plane, which is adopted as the reference plane.
\citet{pok2013} subsequently extended the method to eccentric target orbits in the same reference plane.

Building on these studies, \citet{liang3} and \citet{liang4} extended the method to inclined circular and inclined eccentric target orbits, respectively, removing the requirement that the target orbit lie in the reference plane.
However, \citet{liang4} retained the basic assumptions adopted by \citet{green1982} for the target orbit.
The target's semimajor axis, eccentricity, and inclination are held fixed, while its longitude of the ascending node and argument of periapsis are prescribed to precess at constant rates.
For example, consider a target whose secular evolution is dominated by the central body's oblateness, with the central body's spin axis approximately fixed and its equatorial plane coincident with the perturber's orbital plane.
Taking this common plane as the reference plane, the averaged model to first order in \(J_2\) predicts constant mean semimajor axis, eccentricity, and inclination,
together with uniform precession of the mean longitude of the ascending node and mean argument of periapsis \citep{schaub2001}.

However, this treatment of the target's orbital evolution becomes inadequate when its eccentricity or inclination varies appreciably over secular timescales, or when either precession rate cannot be approximated as constant.
Here, we extend the framework of \citet{liang4}
to calculate the long-term average collision frequency for one such case, in which both the projectile and target undergo LK evolution.

The remainder of this paper is organized as follows.
Section~\ref{p5s2:sec} presents the secular dynamical model and orbital intersection conditions, and describes how the contributions of individual intersection configurations yield single-cycle and long-term average collision frequencies.
Section~\ref{p5s3:sec} analyzes the long-term distribution of starting states of successive projectile LK cycles and examines the dependence of the long-term average collision frequency on initial states.
Section~\ref{p5res:sec} presents the baseline case, results for different initial states, and a comparison with direct N-body simulations.
Section~\ref{p5:conclusion} summarizes the main conclusions.

\section{Calculation of the long-term average collision frequency}\label{p5s2:sec}

Following the overall framework of \citet{liang4}, we extend their method to calculate collision frequencies when both the projectile and target undergo LK secular evolution.
We describe the calculation procedure below to provide a self-contained presentation.
We first determine the times of exact orbital intersection from the secular evolution of both bodies.
For each intersection, we determine the time window from the evolution of both orbits and use it, together with the conditional probability, to evaluate the collision contribution. We then sum these contributions within each projectile LK cycle to calculate the single-cycle collision frequency and average the resulting frequencies over successive cycles to obtain the long-term average collision frequency.

\subsection{Secular dynamical model}\label{p5s2:dyn}

We consider a projectile and a target orbiting the same central body under gravitational perturbations from an external perturber.
The gravitational parameters of the central body and perturber are \(\mu_0\) and \(\mu_P\), respectively, and the perturber moves around the central body on a fixed circular orbit with semimajor axis \(a_P\).
We use a nonrotating Cartesian frame centered on the central body, with the perturber's orbital plane as the reference plane.
The projectile and target are treated as independent perturbed test particles: we neglect their mutual gravitational interaction and their gravitational effects on the central body and perturber.
The subscripts \(p\) and \(T\) denote the projectile and target, respectively, while \(j\) denotes either body.

The mean orbital elements of body \(j\) are denoted by \((a_j,e_j,i_j,\Omega_j,\omega_j)\): the semimajor axis, eccentricity, inclination, longitude of the ascending node, and argument of periapsis, respectively.
To avoid the coordinate singularity associated with the undefined periapsis direction in the circular limit, we introduce the nonsingular variables
\begin{equation}\label{p5s2:kh}
k_j=e_j\cos\omega_j,\qquad
h_j=e_j\sin\omega_j,\qquad
\eta_j=\sqrt{1-k_j^2-h_j^2}=\sqrt{1-e_j^2}.
\end{equation}

We adopt the doubly averaged quadrupole Hamiltonian used by \citet{vok2012} to describe the LK secular evolution of each body.
In this model, the semimajor axis \(a_j\) remains constant, and the Kozai constant is
\begin{equation}\label{p5s2:c}
c_j=\eta_j\cos i_j.
\end{equation}
This conservation law couples the secular variations of eccentricity and inclination.
Using the same dimensionless normalization as in that work, we write the Hamiltonian as
\begin{equation}\label{p5s2:H}
\begin{aligned}
H(k_j,h_j;c_j)
={}&\left[2+3(k_j^2+h_j^2)\right]
\left(\frac{3c_j^2}{\eta_j^2}-1\right)+15(k_j^2-h_j^2)
\left(1-\frac{c_j^2}{\eta_j^2}\right)
=H_j.
\end{aligned}
\end{equation}
The values of \(a_j\), \(c_j\), and \(H_j\) are determined by the initial orbital state of body \(j\).
The pair \((k_j,h_j)\) evolves along the corresponding Hamiltonian level set at fixed \(c_j\).
If this level set contains several mutually disconnected components, the initial orbital state determines which component the trajectory follows and where on that component the motion begins.

The secular evolution equations for body \(j\) can be written in terms of \(k_j,h_j,\Omega_j\) as \citep{kin2007,vok2012}
\begin{equation}\label{p5s2:rates}
\begin{aligned}
\frac{\mathrm d k_j}{\mathrm d t}
&=\frac{3\gamma_j h_j}{4\eta_j^3}
\left[3\eta_j^4-5c_j^2(1-k_j^2)\right],\\
\frac{\mathrm d h_j}{\mathrm d t}
&=\frac{3\gamma_j k_j}{4\eta_j^3}
\left[2\eta_j^4+5c_j^2h_j^2\right],\\
\frac{\mathrm d\Omega_j}{\mathrm d t}
&=-\frac{3\gamma_j c_j}{4\eta_j^2}
\left(1-k_j^2+4h_j^2\right),
\end{aligned}
\end{equation}
where \(\gamma_j= \frac{\mu_P}{a_P^3}\sqrt{\frac{a_j^3}{\mu_0}}\) is a constant coefficient with dimensions of inverse time.

Integrating Eq.~\eqref{p5s2:rates} separately for the two bodies gives their respective \(k_j,h_j,\Omega_j\) as functions of time. These solutions determine the time-dependent mean orbital elements of each body.
The inclination is determined by Eqs.~\eqref{p5s2:kh} and~\eqref{p5s2:c} and does not require a separate integration.
To follow changes in the intersection configurations, we also need the rate of change of inclination.
Differentiating the defining relation for the Kozai constant in Eq.~\eqref{p5s2:c} with respect to time, using its conservation, and substituting Eq.~\eqref{p5s2:rates} gives
\begin{equation}\label{p5s2:idot}
\frac{\mathrm d i_j}{\mathrm d t}
=-\frac{15\gamma_j k_jh_j}{4\eta_j}
\sin i_j\cos i_j.
\end{equation}

The LK period of body \(j\), denoted by \(T_{\mathrm K,j}\), is the shortest positive physical time required for the pair \((k_j,h_j)\) to return to its initial state.\footnote{
This study considers only cases in which the initial orbital state of each body determines a closed Hamiltonian level curve in the \((k_j,h_j)\) plane. Equilibrium points and separatrices are excluded.}
Conservation of \(c_j\) ensures that the inclination also returns to its initial value at that time, whereas the longitude of the ascending node need not return to its initial value after one LK period.

\subsection{Orbital intersection configurations}\label{p5s2:geometry}

At any physical time \(t\), the time-dependent mean orbital elements define instantaneous Keplerian ellipses for the projectile and target.
For any two noncoplanar orbits, every point of intersection lies on the line of intersection of their orbital planes.\footnote{For coplanar orbits, the line of intersection of their orbital planes is not uniquely defined. Therefore, as in \citet{vok2012,pok2013}, this study considers only noncoplanar configurations.}
We can therefore establish exact intersection conditions by comparing the distances from the central body to the points where the two orbits meet this line, separately for each of its two opposite directions.

We define the relative nodal longitude as \(\Delta\Omega=\Omega_p-\Omega_T\).
The mutual inclination \(I\) is the angle between the orbital angular momentum directions of the projectile and target and satisfies
\begin{equation}\label{p5s2:I}
\cos I
=\cos i_p\cos i_T
+\sin i_p\sin i_T\cos\Delta\Omega.
\end{equation}

For noncoplanar configurations (\(0<I<\pi\)), the two orbital planes have a unique line of intersection.
The points where each orbit meets this line are called mutual nodes.
We distinguish the two opposite directions along the line by \(s=\pm1\).
The positive side of the target orbital plane is defined by the direction of the target's orbital angular momentum.
The direction toward the projectile's mutual ascending node relative to the target orbital plane is assigned \(s=+1\); at this node, the projectile crosses from the negative to the positive side of that plane as it moves along its orbit.
The opposite direction is assigned \(s=-1\).

The argument of latitude of the mutual node in direction \(s\) on the orbit of body \(j\) is denoted by \(u_{j,s}\).
This angle is measured from the orbit's ascending node on the reference plane to the mutual node, in the direction of orbital motion.
With the branch convention defined above, the two arguments of latitude can be written as \citep[see][]{liang4}
\begin{equation}\label{p5s2:u}
\begin{aligned}
u_{p,s}
&=\operatorname{atan2}\!\left(
s\sin i_T\sin\Delta\Omega,\,
s\left[
\sin i_p\cos i_T
-\cos i_p\sin i_T\cos\Delta\Omega
\right]\right),\\
u_{T,s}
&=\operatorname{atan2}\!\left(
s\sin i_p\sin\Delta\Omega,\,
s\left[
\sin i_p\cos i_T\cos\Delta\Omega
-\cos i_p\sin i_T
\right]\right).
\end{aligned}
\end{equation}

Let \(r_{j,s}\) be the radial distance from the central body to mutual node \(s\) on the orbit of body \(j\).
The orbit equation can be written in terms of the nonsingular variables defined in Eq.~\eqref{p5s2:kh} as
\begin{equation}\label{p5s2:radius}
r_{j,s}
=\frac{a_j\eta_j^2}
{1+k_j\cos u_{j,s}+h_j\sin u_{j,s}}.
\end{equation}

The two orbits intersect exactly in direction \(s\) if and only if \(r_{p,s}=r_{T,s}\).
Subtracting Eq.~\eqref{p5s2:radius} for the target from that for the projectile gives the signed radial difference between the corresponding mutual nodes:
\begin{equation}\label{p5s2:gap}
\Delta r_s
\equiv r_{p,s}-r_{T,s}
=\frac{a_p}{F_{p,s}F_{T,s}}G_s.
\end{equation}
Here, \(F_{p,s}\) and \(F_{T,s}\) are dimensionless auxiliary quantities defined by \(F_{j,s}=1+k_j\cos u_{j,s}+h_j\sin u_{j,s}\) for \(j=p,T\). The parameter \(\alpha=a_T/a_p\) is the ratio of the target's semimajor axis to that of the projectile. The dimensionless intersection function \(G_s\) is defined by
\begin{equation}\label{p5s2:G}
G_s
=\eta_p^2F_{T,s}
-\alpha\eta_T^2F_{p,s}.
\end{equation}

The exact intersection condition is therefore \(\Delta r_s=0\), which is equivalent to \(G_s=0\).

Substituting the time-dependent mean orbital elements of both bodies into Eq.~\eqref{p5s2:G} gives the intersection function \(G_s(t)\).
Over the time interval under consideration, we solve
\begin{equation}\label{p5s2:root}
G_s(t)=0.
\end{equation}
The roots give the times of exact orbital intersection in the model.
For each branch \(s\), we consider only isolated roots and denote them in chronological order by \(t_{s,w}^{*}\), where \(w\) is the root index.
Each root corresponds to an exact orbital intersection configuration of the two bodies at that time.

\subsection{Collision contributions from intersection configurations}\label{p5s2:contribution}

Because both bodies have finite sizes, collisions remain possible near a time of exact orbital intersection even when the two orbits no longer intersect exactly.
Each exact intersection configuration therefore contributes to the collision probability. This contribution depends on how long the orbits remain sufficiently close near that time and on the conditional probability that the two bodies are in the encounter region at the same time.

The collision contribution \(P_{s,w}\) of each exact intersection configuration is written as the product of two probability factors \citep[see][]{vok2012}:
\begin{equation}\label{p5s2:product}
P_{s,w}=P_{1,s,w}P_{2,s,w}.
\end{equation}
Here, \(P_{1,s,w}\) is the duration for which the two orbits remain sufficiently close near \(t_{s,w}^{*}\), expressed as a fraction of one projectile LK period.
For each passage of the projectile through the encounter region, \(P_{2,s,w}\) is the conditional probability that the target is also in a position allowing a collision, given that the two orbits are sufficiently close.

\paragraph{Probability of orbital proximity.}

The time window around \(t_{s,w}^{*}\) is defined as the continuous time interval during which the local minimum distance between the orbits near the corresponding mutual nodes does not exceed the effective collision radius \(R_{\mathrm{eff}}\).
Within this window, collisions are geometrically possible.
When gravitational focusing is neglected, the effective collision radius is \(R_{\mathrm{eff}}=R_p+R_T\), where \(R_p\) and \(R_T\) are the geometric radii of the projectile and target, respectively.

The quantity \(|\Delta r_s|\) in Eq.~\eqref{p5s2:gap} measures the radial separation of the two mutual nodes, whereas the local minimum distance between the nearby orbital arcs may be smaller.
The orbital arcs are locally approximated by tangent lines at their respective mutual nodes, and the shortest distance between these lines estimates the local minimum distance between the arcs. Using the expression given by \citet{liang4}, the local minimum distance \(D_s\) near the mutual nodes in direction \(s\) is approximated by
\begin{equation}\label{p5s2:D}
D_s\simeq
\frac{\sin I}
{\sqrt{
\sigma_{p,s}^{2}+\sigma_{T,s}^{2}+\sin^{2}I
-2\sigma_{p,s}\sigma_{T,s}\cos I
}}\,|\Delta r_s|.
\end{equation}
Here, \(\sigma_{j,s}\) is shorthand for \((k_j\sin u_{j,s}-h_j\cos u_{j,s})/F_{j,s}\).
Substituting Eq.~\eqref{p5s2:gap} into Eq.~\eqref{p5s2:D} gives
\begin{equation}\label{p5s2:Dcompact}
D_s\simeq\ell_s|G_s|,
\end{equation}
where
\begin{equation}\label{p5s2:ls}
\ell_s=
\frac{a_p\sin I}
{F_{p,s}F_{T,s}
\sqrt{
\sigma_{p,s}^{2}+\sigma_{T,s}^{2}+\sin^{2}I
-2\sigma_{p,s}\sigma_{T,s}\cos I
}}.
\end{equation}

Substituting the time-dependent mean orbital elements of both bodies into Eq.~\eqref{p5s2:Dcompact} gives
\begin{equation}\label{p5s2:Dtime}
D_s(t)\simeq\ell_s(t)|G_s(t)|.
\end{equation}
To estimate the time window, Eq.~\eqref{p5s2:Dtime} is approximated near \(t_{s,w}^{*}\) by replacing \(\ell_s(t)\) with \(\ell_s^{*}\equiv\ell_s(t_{s,w}^{*})\) and expanding \(G_s(t)\) to first order about that time. Since \(G_s(t_{s,w}^{*})=0\), the linearized form of \(G_s(t)\) is
\begin{equation}\label{p5s2:Glin}
G_s(t)\simeq
\left(\frac{\mathrm d G_s}{\mathrm d t}\right)^{*}
(t-t_{s,w}^{*}).
\end{equation}
The derivative \(\mathrm d G_s/\mathrm d t\) is the total derivative with respect to the common physical time \(t\), accounting for the evolution of both bodies' orbital elements.
Here and below, a superscript \(*\) indicates evaluation using the orbital elements of both bodies at the corresponding exact intersection time \(t_{s,w}^{*}\).
Substituting these approximations into Eq.~\eqref{p5s2:Dtime} gives
\begin{equation}\label{p5s2:Dlin}
D_s(t)\simeq
\ell_s^{*}
\left|
\left(\frac{\mathrm d G_s}{\mathrm d t}\right)^{*}
\right|
\,|t-t_{s,w}^{*}|.
\end{equation}
For a root with \(\left(\mathrm d G_s/\mathrm d t\right)^{*}\ne0\), setting the linear approximation in Eq.~\eqref{p5s2:Dlin} equal to \(R_{\mathrm{eff}}\) gives the candidate half-width
\begin{equation}\label{p5s2:halfwindow}
\delta t_{s,w}^{\mathrm{lin}}
=
\frac{R_{\mathrm{eff}}}
{\ell_s^{*}
\left|
\left(\dfrac{\mathrm d G_s}{\mathrm d t}\right)^{*}
\right|}.
\end{equation}
The linear candidate values for the left and right window boundaries, \(t_{s,w}^{\mathrm L}\) and \(t_{s,w}^{\mathrm R}\), are \(t_{s,w}^{*}-\delta t_{s,w}^{\mathrm{lin}}\) and \(t_{s,w}^{*}+\delta t_{s,w}^{\mathrm{lin}}\), respectively.

At exact orbital intersection, \(\left(\eta_p^2F_{T,s}\right)^{*}=\alpha\left(\eta_T^2F_{p,s}\right)^{*}\). Taking the total time derivative of Eq.~\eqref{p5s2:G} and applying this relation gives
\begin{equation}\label{p5s2:Gdot_root}
\begin{aligned}
\left(\frac{\mathrm d G_s}{\mathrm d t}\right)^{*}
=\Bigg\{
\eta_p^2F_{T,s}\Bigg[
&\sigma_{p,s}\frac{\mathrm d u_{p,s}}{\mathrm d t}
-\sigma_{T,s}\frac{\mathrm d u_{T,s}}{\mathrm d t}
+2\left(
\frac{k_T\frac{\mathrm d k_T}{\mathrm d t}
+h_T\frac{\mathrm d h_T}{\mathrm d t}}{\eta_T^2}
-\frac{k_p\frac{\mathrm d k_p}{\mathrm d t}
+h_p\frac{\mathrm d h_p}{\mathrm d t}}{\eta_p^2}
\right)\\
&+\frac{
\frac{\mathrm d k_T}{\mathrm d t}\cos u_{T,s}
+\frac{\mathrm d h_T}{\mathrm d t}\sin u_{T,s}}
{F_{T,s}}
-\frac{
\frac{\mathrm d k_p}{\mathrm d t}\cos u_{p,s}
+\frac{\mathrm d h_p}{\mathrm d t}\sin u_{p,s}}
{F_{p,s}}
\Bigg]
\Bigg\}^{*}.
\end{aligned}
\end{equation}
The time derivatives of the arguments of latitude follow by taking the total time derivatives of Eq.~\eqref{p5s2:u}:
\begin{equation}\label{p5s2:udot}
\begin{aligned}
\frac{\mathrm d u_{p,s}}{\mathrm d t}
={}&
\frac{\sin\Delta\Omega
\left(
\sin i_p\,\frac{\mathrm d i_T}{\mathrm d t}
-\sin i_T\cos I\,\frac{\mathrm d i_p}{\mathrm d t}
\right)}
{\sin^2 I}\\
&+
\frac{\sin i_T
\left(
\sin i_p\cos i_T\cos\Delta\Omega
-\cos i_p\sin i_T
\right)}
{\sin^2 I}
\left(
\frac{\mathrm d\Omega_p}{\mathrm d t}
-\frac{\mathrm d\Omega_T}{\mathrm d t}
\right),\\[2mm]
\frac{\mathrm d u_{T,s}}{\mathrm d t}
={}&
\frac{\sin\Delta\Omega
\left(
\sin i_p\cos I\,\frac{\mathrm d i_T}{\mathrm d t}
-\sin i_T\,\frac{\mathrm d i_p}{\mathrm d t}
\right)}
{\sin^2 I}\\
&+
\frac{\sin i_p
\left(
\sin i_p\cos i_T
-\cos i_p\sin i_T\cos\Delta\Omega
\right)}
{\sin^2 I}
\left(
\frac{\mathrm d\Omega_p}{\mathrm d t}
-\frac{\mathrm d\Omega_T}{\mathrm d t}
\right).
\end{aligned}
\end{equation}
In Eqs.~\eqref{p5s2:Gdot_root} and~\eqref{p5s2:udot}, the derivatives \(\mathrm d k_j/\mathrm d t\), \(\mathrm d h_j/\mathrm d t\), and \(\mathrm d\Omega_j/\mathrm d t\) are given by Eq.~\eqref{p5s2:rates}, and \(\mathrm d i_j/\mathrm d t\) is given by Eq.~\eqref{p5s2:idot}.
Thus, the duration of the time window depends on the orbital evolution of both bodies near the exact intersection time.

Each candidate boundary is checked by evaluating the local minimum distance using the tangent approximation in Eq.~\eqref{p5s2:Dtime}.
A candidate boundary is accepted as the final boundary on that side if \(|D_s(t)-R_{\mathrm{eff}}|/R_{\mathrm{eff}}\leq\varepsilon_{\mathrm{win}}\), where \(\varepsilon_{\mathrm{win}}\) is the relative tolerance for this distance check.
If the tolerance is not met on one side, a search is performed from the exact intersection time in the corresponding time direction, and the first solution of \(\ell_s(t)|G_s(t)|=R_{\mathrm{eff}}\) is adopted as the final boundary on that side.
For an isolated root with \(\left(\mathrm d G_s/\mathrm d t\right)^{*}=0\), the calculation of linear candidate boundaries is skipped. Both final window boundaries are obtained directly by the search described above, applied separately in the two time directions from the root.

Dividing the full duration between the final window boundaries \(t_{s,w}^{\mathrm L}\) and \(t_{s,w}^{\mathrm R}\) by the projectile LK period gives
\begin{equation}\label{p5s2:P1}
P_{1,s,w}
=
\frac{t_{s,w}^{\mathrm R}-t_{s,w}^{\mathrm L}}
{T_{\mathrm K,p}}.
\end{equation}

\paragraph{Conditional probability.}

The conditional probability derived by \citet{green1982} accounts for simultaneous presence of both bodies in the encounter region, given sufficiently close orbits.
Its derivation uses uniform rectilinear motion with Keplerian velocities at the exact intersection point, a uniformly distributed target mean anomaly, and uniform averaging over \(0\leq D_s\leq R_{\mathrm{eff}}\).
The equivalent expression given by \citet{liang4} is used here:
\begin{equation}\label{p5s2:P2}
P_{2,s,w}
=
\left\{
\frac{R_{\mathrm{eff}}\eta_T}{4a_TF_{T,s}}
\left[
\frac{
1+\sigma_{p,s}^{2}
+\alpha\frac{\eta_T^{2}}{\eta_p^{2}}
(1+\sigma_{T,s}^{2})
-2\sqrt{\alpha}\frac{\eta_T}{\eta_p}
(\sigma_{p,s}\sigma_{T,s}+\cos I)
}{
(1+\sigma_{p,s}^{2})(1+\sigma_{T,s}^{2})
-(\sigma_{p,s}\sigma_{T,s}+\cos I)^{2}
}
\right]^{1/2}
\right\}^{*}.
\end{equation}
When the two local orbital tangents at the exact orbital intersection point are nearly parallel or antiparallel, this approximation may yield \(P_{2,s,w}>1\).
The probability is therefore capped at unity by taking \(P_{2,s,w}=\min\{P_{2,s,w},1\}\).

\subsection{Single-cycle collision frequency and long-term averaging}\label{p5s2:single_cycle}

We obtain the single-cycle collision frequency by summing the contributions from all exact intersection configurations that occur during one projectile LK cycle and converting the sum to a frequency.
Starting at \(t_0\), we divide physical time into analysis intervals of duration \(T_{\mathrm K,p}\), with boundaries
\begin{equation}\label{p5s2:cycle_boundaries}
t_n=t_0+nT_{\mathrm K,p},
\qquad n=0,1,2,\ldots .
\end{equation}
The single-cycle collision frequency \(\Gamma^{(n)}\) for the \(n\)th analysis interval, \([t_{n-1},t_n)\), is
\begin{equation}\label{p5s2:single_cycle_rate}
\Gamma^{(n)}
=
\frac{1}{T_{\mathrm{orb},p}}
\sum_{s=\pm1}
\sum_{\substack{w\\
t_{s,w}^{*}\in[t_{n-1},t_n)}}
P_{s,w}.
\end{equation}
Here, \(T_{\mathrm{orb},p}=2\pi\sqrt{a_p^3/\mu_0}\) is the projectile's Keplerian orbital period.
Each intersection configuration contributes to the sum for the analysis interval containing its exact intersection time.
If a time window extends across an analysis interval boundary, its contribution is still calculated using the full window duration.

For the same physical system, the single-cycle collision frequency for each analysis interval is determined by the starting state \((k_p,h_p,k_T,h_T,\Delta\Omega)\) of that analysis interval.
These states generally differ between analysis intervals, so the intersection configurations encountered and their collision contributions can also differ.

To combine the collision contributions from the different intersection configurations encountered over successive analysis intervals, the cumulative mean collision frequency is defined as the arithmetic mean of the single-cycle collision frequencies over the first \(N\) analysis intervals:
\begin{equation}\label{p5s2:cumulative_average}
\Gamma_{(N)}
=
\frac{1}{N}\sum_{n=1}^{N}\Gamma^{(n)}.
\end{equation}
All analysis intervals have the same duration \(T_{\mathrm K,p}\) and therefore carry equal time weights in this average. The long-term average collision frequency is defined as \(\Gamma=\lim_{N\to\infty}\Gamma_{(N)}\).

To estimate this limit numerically, the cumulative mean \(\Gamma_{(N)}\) is first calculated using \(N_{\min}\) analysis intervals and then updated as the cumulative number of analysis intervals \(N\) is successively doubled.
The numerical stability of the cumulative mean is assessed from the relative change between successive cumulative means using the criterion
\begin{equation}\label{p5s2:stability_criterion}
\frac{|\Gamma_{(2N)}-\Gamma_{(N)}|}
{\Gamma_{(2N)}}<\varepsilon,
\qquad \Gamma_{(2N)}>0,
\end{equation}
where \(\varepsilon>0\) is a prescribed threshold for the relative change.
Once this criterion has been satisfied in \(N_{\mathrm{pass}}\) consecutive checks, the final cumulative mean is adopted as a numerical estimate of the long-term average collision frequency.
If \(N\) reaches the prescribed upper limit \(N_{\max}\) before the stopping condition is met, the cumulative mean at that limit is reported, together with an explicit statement that the stopping condition was not satisfied.

\section{Dependence of the long-term average collision frequency on initial states}\label{p5s3:sec}

This section examines whether the long-term average collision frequency depends on the initial states of the two bodies,
with their semimajor axes, Kozai constants, and respective closed Hamiltonian level curves held fixed.

\subsection{Starting state of each analysis interval}\label{p5s3:distribution}

The single-cycle collision frequency is determined by the five quantities \((k_p,h_p,k_T,h_T,\Delta\Omega)\) that specify the starting state of each analysis interval.
According to Eq.~\eqref{p5s2:cycle_boundaries}, successive analysis interval starts are separated by \(T_{\mathrm K,p}\), so the projectile's \(k_p,h_p\) have the same values at every analysis interval start:
\begin{equation}\label{p5s3:projectile_return}
k_p(t_{n-1})=k_p(t_0),
\qquad
h_p(t_{n-1})=h_p(t_0).
\end{equation}
The starting state of each analysis interval can therefore be specified by the three quantities \((k_T,h_T,\Delta\Omega)\).

The target's \(k_T\) and \(h_T\) evolve together along the same closed Hamiltonian level curve and repeat the same evolution in each LK cycle.
Both quantities can therefore be specified jointly by the target's time within that cycle.

A fixed state \((k_T^{\mathrm{ref}},h_T^{\mathrm{ref}})\) is selected on the target's closed Hamiltonian level curve as the reference state.
The relative time within the target's LK cycle is defined as the physical time elapsed since the target most recently reached this reference state.
The reference time \(t^{\mathrm{ref}}_T\) is defined as the time at which the target first reaches the reference state as it evolves from its initial state \((k_T(t_0),h_T(t_0))\) at \(t_0\).
The relative time within the cycle at physical time \(t\) is then given by
\begin{equation}\label{p5s3:target_position}
t_T(t)
=
(t-t^{\mathrm{ref}}_T)\bmod T_{\mathrm K,T},
\qquad
0\leq t_T(t)<T_{\mathrm K,T}.
\end{equation}
The starting state of each analysis interval can thus be specified more compactly
by the two quantities \(t_T(t_{n-1})\) and \(\Delta\Omega(t_{n-1})\).

\subsection{Evolution of the starting state of each analysis interval}\label{p5s3:mapping}

According to Eq.~\eqref{p5s2:cycle_boundaries}, the start times of successive analysis intervals are separated by \(T_{\mathrm K,p}\), so the relative time within the target's LK cycle satisfies
\begin{equation}\label{p5s3:target_recurrence}
t_T(t_n)
=
\bigl[t_T(t_{n-1})+T_{\mathrm K,p}\bigr]
\bmod T_{\mathrm K,T}.
\end{equation}

The relative nodal longitude \(\Delta\Omega\), the other quantity specifying the starting state, evolves through the nodal precession of both bodies.
Equation~\eqref{p5s2:rates} shows that each body's nodal precession rate varies periodically as its \(k_j,h_j\) evolve.
Averaging this periodic rate over one LK cycle gives the mean nodal precession rate of body \(j\):
\begin{equation}\label{p5s3:mean_node_rate}
\overline{\frac{\mathrm d\Omega_j}{\mathrm d t}}
=
\frac{1}{T_{\mathrm K,j}}
\int_{t_0}^{t_0+T_{\mathrm K,j}}
\frac{\mathrm d\Omega_j}{\mathrm d t}\,\mathrm d t,
\qquad j=p,T.
\end{equation}
For an integration interval of duration \(T_{\mathrm K,j}\), the average in Eq.~\eqref{p5s3:mean_node_rate} is independent of the integration start time.
The longitude of the ascending node \(\Omega_j\) of body \(j\), treated as a continuously accumulated angle, can therefore be decomposed into a linear precession term and a periodic correction:
\begin{equation}\label{p5s3:node_decomposition}
\Omega_j(t)
=
\overline{\Omega}_j(t)
+
\widetilde{\Omega}_j(t).
\end{equation}
The linear precession term \(\overline{\Omega}_j(t)\) evolves at the mean nodal precession rate,
\begin{equation}\label{p5s3:linear_node}
\overline{\Omega}_j(t)
=
\Omega_j(t_0)
+
\overline{\frac{\mathrm d\Omega_j}{\mathrm d t}}\,
(t-t_0),
\end{equation}
and the periodic correction \(\widetilde{\Omega}_j(t)\) satisfies \(\widetilde{\Omega}_j(t_0)=0\).
Over any interval of duration \(T_{\mathrm K,j}\), both \(\Omega_j(t)\) and \(\overline{\Omega}_j(t)\) change by \(\overline{\mathrm d\Omega_j/\mathrm d t}\,T_{\mathrm K,j}\).
The periodic correction \(\widetilde{\Omega}_j(t)\) therefore has period \(T_{\mathrm K,j}\):
\begin{equation}\label{p5s3:node_correction_periodicity}
\widetilde{\Omega}_j(t+T_{\mathrm K,j})
=
\widetilde{\Omega}_j(t).
\end{equation}

The relative nodal longitude can be expressed by applying Eq.~\eqref{p5s3:node_decomposition} to both bodies:
\begin{equation}\label{p5s3:relative_node_decomposition}
\begin{aligned}
\Delta\Omega(t)
=
\bigl[\overline{\Omega}_p(t)-\overline{\Omega}_T(t)\bigr]
+
\bigl[\widetilde{\Omega}_p(t)-\widetilde{\Omega}_T(t)\bigr].
\end{aligned}
\end{equation}
Here, the first term on the right-hand side is the linear term \(\overline{\Delta\Omega}(t)\), and the second is the correction \(\widetilde{\Delta\Omega}(t)\).
Subtracting the target's linear precession term from the projectile's using Eq.~\eqref{p5s3:linear_node} yields
\begin{equation}\label{p5s3:linear_relative_node}
\begin{aligned}
\overline{\Delta\Omega}(t)
&=
\Delta\Omega(t_0)
+
\left(
\overline{\frac{\mathrm d\Omega_p}{\mathrm d t}}
-
\overline{\frac{\mathrm d\Omega_T}{\mathrm d t}}
\right)(t-t_0).
\end{aligned}
\end{equation}
Although each individual correction \(\widetilde{\Omega}_j(t)\) is periodic, the two bodies' LK periods \(T_{\mathrm K,p}\) and \(T_{\mathrm K,T}\) may differ.
The correction \(\widetilde{\Delta\Omega}(t)\) is therefore not necessarily periodic.

Evaluating Eq.~\eqref{p5s3:linear_relative_node} at \(t_n\) and \(t_{n-1}\), subtracting, and reducing modulo \(2\pi\) gives
\begin{equation}\label{p5s3:linear_relative_node_return}
\begin{aligned}
\overline{\Delta\Omega}(t_n)\bmod 2\pi
={}
\Biggl[
\overline{\Delta\Omega}(t_{n-1})
+
\left(
\overline{\frac{\mathrm d\Omega_p}{\mathrm d t}}
-
\overline{\frac{\mathrm d\Omega_T}{\mathrm d t}}
\right)T_{\mathrm K,p}
\Biggr]\bmod 2\pi.
\end{aligned}
\end{equation}
The second term inside the brackets is the fixed increment of the linear term over one analysis interval and is independent of the analysis interval index \(n\).

\subsection{Long-term distribution of the starting-state point sequence}\label{p5s3:long_term_distribution}

The recurrences in Eqs.~\eqref{p5s3:target_recurrence} and~\eqref{p5s3:linear_relative_node_return} determine the values of \(t_T(t_{n-1})\) and \(\overline{\Delta\Omega}(t_{n-1})\bmod 2\pi\) at the starts of successive analysis intervals.
These pairs form a sequence of starting-state points in the two-dimensional parameter region \([0,T_{\mathrm K,T})\times[0,2\pi)\).
Its long-term distribution within this region can be classified according to the integer relations satisfied by the normalized increments of the two recurrences.
These dimensionless increments are obtained by normalizing the fixed increments in Eqs.~\eqref{p5s3:target_recurrence} and~\eqref{p5s3:linear_relative_node_return} by \(T_{\mathrm K,T}\) and \(2\pi\), respectively:
\begin{equation}\label{p5s3:normalized_increments}
\begin{aligned}
Z_T &= \frac{T_{\mathrm K,p}}{T_{\mathrm K,T}},\\
Z_\Omega &= \frac{T_{\mathrm K,p}}{2\pi}
\left(
\overline{\frac{\mathrm d\Omega_p}{\mathrm d t}}
-
\overline{\frac{\mathrm d\Omega_T}{\mathrm d t}}
\right).
\end{aligned}
\end{equation}
The quantity \(Z_T\) is the ratio of the analysis interval duration, namely the projectile LK period, to the target LK period.
The quantity \(Z_\Omega\) is the linear increment of the relative nodal longitude over one analysis interval, expressed in units of \(2\pi\).

If there exist integers \(q_0,q_1,q_2\), with \(q_1,q_2\) not both zero, such that
\begin{equation}\label{p5s3:joint_commensurability}
q_1Z_T+q_2Z_\Omega=q_0,
\end{equation}
then \(1,Z_T,Z_\Omega\) are linearly dependent over the rationals, corresponding to the rationally dependent case.
If no such integer relation exists, \(1,Z_T,Z_\Omega\) are linearly independent over the rationals,
corresponding to the rationally independent case.

\begin{figure}[!b]
\centering
\includegraphics[width=\linewidth]{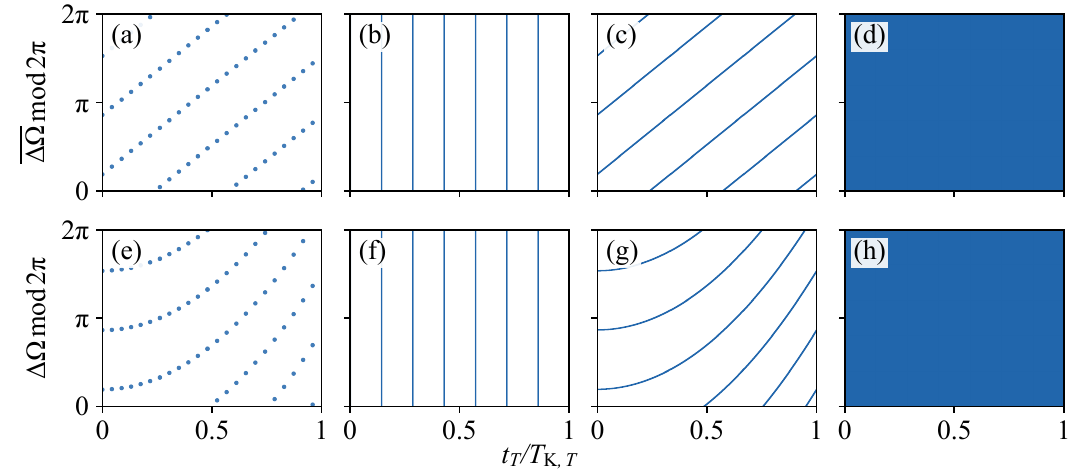}
\caption{
Distributions of the starting-state point sequences of successive analysis intervals in the parameter region before and after an angular mapping.
Panels (a)--(d) show the distributions of the starting-state point sequences of successive analysis intervals in the \((t_T,\overline{\Delta\Omega}\bmod 2\pi)\) parameter region. The prescribed \((Z_T,Z_\Omega)\) values and integer relations are:
(a) \((4/23,35/69)\), with \(23Z_T+69Z_\Omega=39\);
(b) \((2/7,\sqrt{2})\), with \(7Z_T+0Z_\Omega=2\);
(c) \((\sqrt{2},\sqrt{2}+1/3)\), with \(-3Z_T+3Z_\Omega=1\);
(d) \((\sqrt{2},\sqrt{3})\), with no nontrivial integer relation of the form in Eq.~\eqref{p5s3:joint_commensurability}.
Panels (e)--(h) show the distributions of the same starting-state point sequences as in panels (a)--(d), respectively, after mapping from the \((t_T,\overline{\Delta\Omega}\bmod 2\pi)\) parameter region to the \((t_T,\Delta\Omega\bmod 2\pi)\) parameter region.
The prescribed mapping is \(\Delta\Omega=[\overline{\Delta\Omega}-2\pi(t_T/T_{\mathrm K,T})(1-t_T/T_{\mathrm K,T})]\bmod 2\pi\).
For each column, the recurrences in Eqs.~\eqref{p5s3:target_recurrence} and~\eqref{p5s3:linear_relative_node_return} generate \(10^6\) points from the initial point \(\bigl(t_T/T_{\mathrm K,T},\overline{\Delta\Omega}\bigr)=(0,\pi/5)\).
This figure is a mathematical illustration.
}
\label{fig:sampling_distributions}
\end{figure}

These two cases yield four distribution patterns for the starting-state point sequence within the parameter region: three in the rationally dependent case and one in the rationally independent case.
\begin{enumerate}
\renewcommand{\labelenumi}{(\alph{enumi})}
\item Rationally dependent case with \(Z_T\) and \(Z_\Omega\) both rational: the sequence repeats periodically among a finite number of points.
\item Rationally dependent case with exactly one of \(Z_T\) and \(Z_\Omega\) rational: the points lie on a finite number of line segments parallel to one coordinate axis.
\item Rationally dependent case with \(Z_T\) and \(Z_\Omega\) both irrational: the points lie on a finite number of parallel oblique line segments.
\item Rationally independent case with \(Z_T\) and \(Z_\Omega\) both irrational:
the points are uniformly distributed throughout the parameter region.
\end{enumerate}
Panels (a)--(d) of Fig.~\ref{fig:sampling_distributions} show the starting-state point sequences in the \((t_T,\overline{\Delta\Omega}\bmod 2\pi)\) parameter region, in the order listed above.

The periodicity of the periodic correction \(\widetilde{\Omega}_p(t)\) [Eq.~\eqref{p5s3:node_correction_periodicity}], together with the initial condition \(\widetilde{\Omega}_p(t_0)=0\), implies \(\widetilde{\Omega}_p(t_{n-1})=0\).
Evaluation of Eq.~\eqref{p5s3:relative_node_decomposition} at \(t=t_{n-1}\) using this result, followed by reduction of both sides modulo \(2\pi\), yields
\begin{equation}\label{p5s3:actual_relative_node}
\Delta\Omega(t_{n-1})\bmod 2\pi
=
\left[
\overline{\Delta\Omega}(t_{n-1})
-
\widetilde{\Omega}_T(t_{n-1})
\right]\bmod 2\pi.
\end{equation}
At the start of each analysis interval, Eq.~\eqref{p5s3:actual_relative_node} maps the linear term \(\overline{\Delta\Omega}(t_{n-1})\)
to the actual relative nodal longitude \(\Delta\Omega(t_{n-1})\) by an angular shift of \(-\widetilde{\Omega}_T(t_{n-1})\), modulo \(2\pi\).
The periodicity of \(\widetilde{\Omega}_T(t)\) [Eq.~\eqref{p5s3:node_correction_periodicity}] implies that this shift depends only on the target's relative time \(t_T(t_{n-1})\) within its LK cycle.
Moreover, the shift is continuous in \(t_T\), with equal values at \(t_T=0\) and \(t_T=T_{\mathrm K,T}\).
Since \(t_T\) remains unchanged, the mapping from \((t_T,\overline{\Delta\Omega}\bmod 2\pi)\) to \((t_T,\Delta\Omega\bmod 2\pi)\) preserves area within the parameter region.

Panels (e)--(h) of Fig.~\ref{fig:sampling_distributions} show the corresponding starting-state point sequences mapped into the \((t_T,\Delta\Omega\bmod 2\pi)\) parameter region using a prescribed angular shift.
In the rationally dependent cases, distributions along straight line segments may be mapped to distributions along curves.
In the rationally independent case, the points remain uniformly distributed throughout the parameter region after mapping.

\subsection{Independence from initial states}\label{p5s3:independence}

In the dimensionless parameter plane, the parameter pairs \((Z_T,Z_\Omega)\) satisfying at least one integer relation of the form in Eq.~\eqref{p5s3:joint_commensurability} form a set of measure zero.
Equivalently, this set has zero area, so almost all parameter pairs \((Z_T,Z_\Omega)\) in this plane correspond to the rationally independent case.
Accordingly, the following analysis focuses on the dependence of the long-term average collision frequency on initial states in this case.

The initial time \(t_0\), the projectile's initial state \((k_p(t_0),h_p(t_0),\Omega_p(t_0))\), and the selected target reference state \((k_T^{\mathrm{ref}},h_T^{\mathrm{ref}})\) are held fixed.
Changing the target's initial position on its original closed Hamiltonian level curve changes its relative time within the LK cycle at \(t_0\), \(t_T(t_0)\), and hence the horizontal coordinate of the first starting-state point.
Changing the target's initial longitude of the ascending node \(\Omega_T(t_0)\) changes the initial relative nodal longitude \(\Delta\Omega(t_0)\), and hence the vertical coordinate of that point.

Changing only the target's initial position on this curve and its initial longitude of the ascending node leaves \(Z_T\) and \(Z_\Omega\) unchanged.
Thus, \(1,Z_T,Z_\Omega\) remain linearly independent over the rationals.
Regardless of the position of the first starting-state point, the recurrences in Eqs.~\eqref{p5s3:target_recurrence} and~\eqref{p5s3:linear_relative_node_return} therefore generate a starting-state point sequence uniformly distributed throughout the \((t_T,\overline{\Delta\Omega}\bmod 2\pi)\) parameter region \([0,T_{\mathrm K,T})\times[0,2\pi)\).

Each starting-state point in the \((t_T,\Delta\Omega\bmod 2\pi)\) parameter region corresponds to a single-cycle collision frequency.
The arithmetic mean of these frequencies over the first \(N\) analysis intervals is the cumulative mean collision frequency \(\Gamma_{(N)}\), defined in Eq.~\eqref{p5s2:cumulative_average}.
As \(N\) increases, the long-term distribution of the starting-state point sequence determines the relative weights of different states in the long-term average.
The common uniform long-term distribution assigns the same statistical weights to the single-cycle collision frequencies associated with different starting states, yielding the same long-term average collision frequency \(\Gamma\) for different target initial states.

The same argument applies after exchanging the projectile and target labels within the same physical problem.
That is, changing the initial position of the newly designated target, namely the original projectile, along its original closed Hamiltonian level curve or changing its initial longitude of the ascending node also leaves the long-term average collision frequency unchanged.
In summary, when \(1,Z_T,Z_\Omega\) are linearly independent over the rationals and the semimajor axes, Kozai constants, and original closed Hamiltonian level curves of both bodies remain fixed, the long-term average collision frequency is independent of their initial positions along those curves and their initial longitudes of the ascending nodes.

\section{Numerical results}\label{p5res:sec}

Numerical tests assess the stability of the cumulative mean collision frequency and its dependence on initial states.
The resulting survival prediction is also compared with direct N-body simulations using REBOUND.

\subsection{Baseline case and numerical stability}\label{p5res:baseline}

In the baseline case, Case~0, Earth is the central body and the Sun is the external perturber.
The Sun--Earth relative orbit is circular, and both the target and projectile are massless test particles orbiting Earth.
Table~\ref{p5res:tab:physical} lists their initial orbital elements and the physical and numerical parameters.

\begin{table}[htbp]
\centering
\caption{Parameters for the baseline case, Case~0.}
\label{p5res:tab:physical}
\small
\renewcommand{\arraystretch}{1.08}
\setlength{\tabcolsep}{12pt}
\begin{tabular*}{0.60\linewidth}{@{\extracolsep{\fill}}ccc@{}}
\toprule
Parameter & Target \(T\) & Projectile \(p\) \\
\midrule
\multicolumn{3}{c}{\textbf{Initial conditions}} \\
\addlinespace[2pt]
\(a_j\,[\mathrm{au}]\)       & 0.0020 & 0.0022 \\
\(e_j\)                     & 0.30   & 0.25   \\
\(i_j\,[{}^\circ]\)         & 75     & 80     \\
\(\omega_j\,[{}^\circ]\)    & 0      & 90     \\
\(\Omega_j\,[{}^\circ]\)    & 0      & 20     \\

\addlinespace[7pt]
\multicolumn{3}{c}{\textbf{Physical parameters}} \\
\addlinespace[2pt]
\(R_j\,[\mathrm{km}]\)
& 50 & 50 \\
\(\mu_0\,[\mathrm{m^3\,s^{-2}}]\)
& \multicolumn{2}{c}{\(3.9860\times10^{14}\)} \\
\(\mu_P\,[\mathrm{m^3\,s^{-2}}]\)
& \multicolumn{2}{c}{\(1.3271\times10^{20}\)} \\
\(a_P\,[\mathrm{au}]\)
& \multicolumn{2}{c}{1} \\

\addlinespace[7pt]
\multicolumn{3}{c}{\textbf{Numerical parameters}} \\
\addlinespace[2pt]
\(N_{\min}\)
& \multicolumn{2}{c}{\(2^{10}\)} \\
\(N_{\max}\)
& \multicolumn{2}{c}{\(2^{20}\)} \\
\(\varepsilon_{\mathrm{win}}\)
& \multicolumn{2}{c}{\(1\%\)} \\
\(\varepsilon\)
& \multicolumn{2}{c}{\(0.2\%\)} \\
\(N_{\mathrm{pass}}\)
& \multicolumn{2}{c}{2} \\
\bottomrule
\end{tabular*}

\par\smallskip
\begin{minipage}{0.96\linewidth}
\end{minipage}
\end{table}

\begin{figure}[tbp]
\centering
\includegraphics[width=\linewidth]{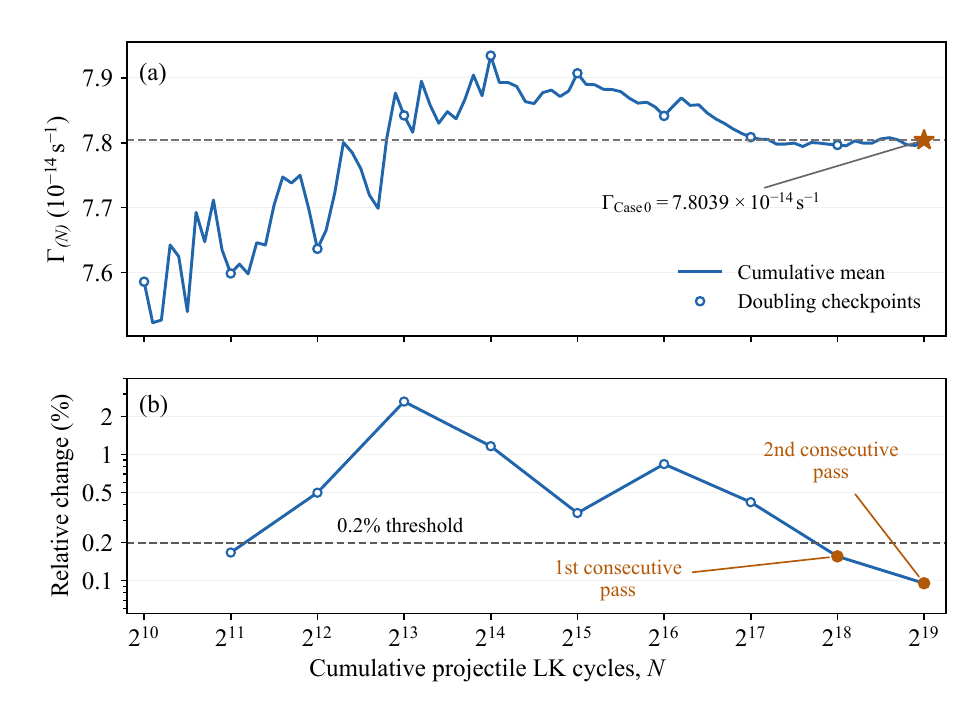}
\caption{
Numerical stability and stopping checks for Case~0.
(a) The blue curve shows \(\Gamma_{(N)}\), with open circles marking checkpoints at which the cumulative number of analysis intervals \(N\) is doubled.
The horizontal dashed line marks \(\Gamma_{\mathrm{Case}\,0}\), and the orange star marks the final estimate.
(b) The relative change \(\lvert\Gamma_{(N)}-\Gamma_{(N/2)}\rvert/\Gamma_{(N)}\) is expressed as a percentage, with the dashed line marking the \(0.2\%\) threshold.
The orange points at \(N=2^{18}\) and \(2^{19}\) mark the two consecutive passes.
}
\label{p5res:fig:convergence}
\end{figure}

Figure~\ref{p5res:fig:convergence} shows the cumulative mean collision frequency \(\Gamma_{(N)}\) and its relative change as functions of the cumulative number of analysis intervals \(N\).
The cumulative mean fluctuates markedly at first and shows much smaller fluctuations as \(N\) increases.

At \(N=2^{19}\), the relative change satisfied Eq.~\eqref{p5s2:stability_criterion} for the second consecutive check, meeting the stopping condition.
The resulting cumulative mean was \(\Gamma_{(2^{19})}=7.8039\times10^{-14}\,\mathrm{s}^{-1}\).
This value is adopted as the Case~0 estimate of the long-term average collision frequency, denoted by \(\Gamma_{\mathrm{Case}\,0}\).

\subsection{Collision frequencies for different initial states}\label{p5res:initial_states}

This subsection tests the theoretical prediction in Section~\ref{p5s3:independence} through numerical cases constructed by changing the initial orbital states of both the projectile and target relative to Case~0.

In practice, finite numerical precision in the LK periods and mean nodal precession rates makes it difficult to establish rigorously whether \(1,Z_T,Z_\Omega\) are linearly independent over the rationals.
The Case~0 starting-state point sequence is therefore examined for approximate uniformity over the finite number of analysis intervals considered.

For Case~0, the LK periods and mean nodal precession rates give \(Z_T\simeq0.42710123\) and \(Z_\Omega\simeq-0.08260299\).
The target's initial state \((k_T(t_0),h_T(t_0))\) in Case~0 is used as the fixed reference state \((k_T^{\mathrm{ref}},h_T^{\mathrm{ref}})=(0.30,0)\), corresponding to \(t_T(t_0)=0\).

\begin{figure}[tbp]
\centering
\includegraphics[width=0.80\linewidth]{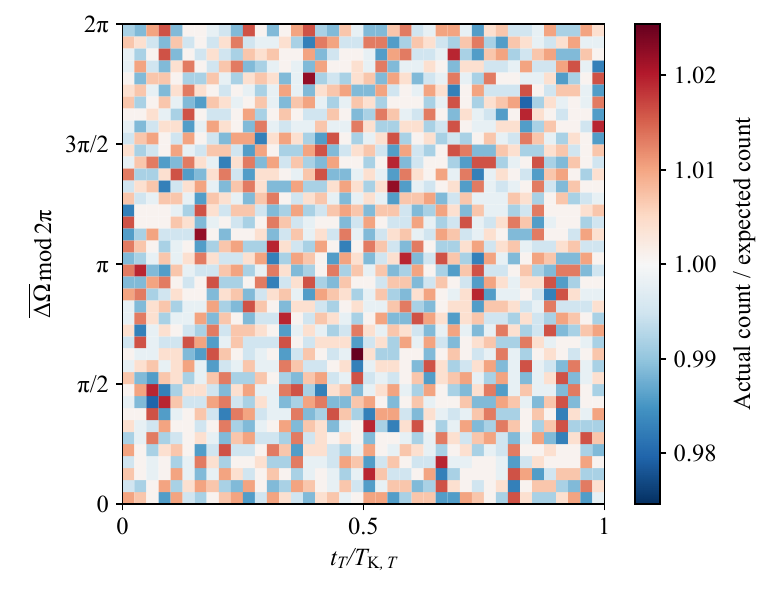}
\caption{
Distribution of the Case~0 starting-state point sequence in the \((t_T,\overline{\Delta\Omega}\bmod 2\pi)\) parameter region.
}
\label{p5res:fig:joint_distribution}
\end{figure}

With \(\bigl(t_T(t_0),\overline{\Delta\Omega}(t_0)\bmod 2\pi\bigr)=(0,\pi/9)\) as the initial point, the first \(N=2^{19}\) starting-state points in the sequence are generated using the recurrences in Eqs.~\eqref{p5s3:target_recurrence} and~\eqref{p5s3:linear_relative_node_return}.
The parameter region \([0,T_{\mathrm K,T})\times[0,2\pi)\) is divided into a uniform \(40\times40\) grid, and the points in each cell are counted.
For each cell, the ratio of the actual point count to the expected count for a uniform distribution is calculated.
The resulting ratios in the \((t_T,\overline{\Delta\Omega}\bmod 2\pi)\) parameter region are shown in Fig.~\ref{p5res:fig:joint_distribution}.
The relative standard deviation of these ratios is \(0.835\%\), indicating an approximately uniform point distribution at the sample size and grid resolution considered.

With Case~0 as the template, eight new numerical cases are constructed by changing the initial orbital states of the projectile and target.
These cases, denoted by Case~\(\nu\) with \(\nu=1,\ldots,8\), differ from Case~0 only in the initial states of the two bodies.
For Case~\(\nu\), the projectile and target are evolved from their respective Case~0 initial states for physical times \(\nu T_{\mathrm K,p}/8\) and \(\nu T_{\mathrm K,T}/8\), respectively.
The resulting \((k_p,h_p)\) and \((k_T,h_T)\) are assigned as their initial values at \(t_0\).
The initial longitudes of the ascending nodes \(\Omega_p(t_0)\) and \(\Omega_T(t_0)\) are drawn independently from a uniform distribution on \([0,2\pi)\).
Because \(T_{\mathrm K,p}\ne T_{\mathrm K,T}\), the forward evolution used to construct the new initial conditions spans different physical durations for the two bodies.
Both bodies' semimajor axes, Kozai constants, and original closed Hamiltonian level curves remain fixed.
All other physical and numerical parameters and the collision frequency calculation method are identical to Case~0.

\begin{table}[htbp]
\centering
\caption{Cumulative mean collision frequencies for different initial states.}
\label{p5res:tab:rates}
\begin{tabular}{lcccc}
\toprule
Case & \(N\) & \(\Gamma_{\mathrm{Case}\,\nu}\,[10^{-14}\,\mathrm{s}^{-1}]\)
& \shortstack{Stopping condition\\met} & \(\delta_\nu\,[\%]\) \\
\midrule
Case~1 & \(2^{20}\) & 7.8083 & No  & \(+0.0563\) \\
Case~2 & \(2^{20}\) & 7.8194 & No  & \(+0.1996\) \\
Case~3 & \(2^{19}\) & 7.8287 & Yes & \(+0.3181\) \\
Case~4 & \(2^{19}\) & 7.8104 & Yes & \(+0.0843\) \\
Case~5 & \(2^{18}\) & 7.8057 & Yes & \(+0.0237\) \\
Case~6 & \(2^{20}\) & 7.8095 & No  & \(+0.0724\) \\
Case~7 & \(2^{20}\) & 7.8000 & Yes & \(-0.0496\) \\
Case~8 & \(2^{20}\) & 7.8211 & Yes & \(+0.2214\) \\
\bottomrule
\end{tabular}

\par\smallskip
\begin{minipage}{0.96\linewidth}
\footnotesize
The quantity \(N\) is the cumulative number of analysis intervals at the end of each calculation, and \(\Gamma_{\mathrm{Case}\,\nu}\) is the corresponding cumulative mean \(\Gamma_{(N)}\).
The relative difference from Case~0 is \(\delta_\nu=\frac{\Gamma_{\mathrm{Case}\,\nu}-\Gamma_{\mathrm{Case}\,0}}{\Gamma_{\mathrm{Case}\,0}}\times100\%\).
\end{minipage}
\end{table}

Table~\ref{p5res:tab:rates} summarizes the numerical results for Cases~1--8.
Cases~1, 2, and~6 reached \(N_{\max}=2^{20}\) without satisfying the stopping condition, and their reported collision frequencies are the cumulative means at this limit.
The remaining five cases satisfied the stopping condition.

The cumulative mean collision frequencies for the eight cases differ from the Case~0 estimate by \(-0.0496\%\) to \(+0.3181\%\).
These results provide numerical support for the predicted independence of the long-term average collision frequency from the initial states.

\subsection{Comparison with direct N-body simulations using REBOUND}\label{p5res:rebound}

Direct simulations used the open-source N-body package REBOUND \citep{rein2012} to compare the projectile survival curve with the semi-analytical prediction based on \(\Gamma_{\mathrm{Case}\,0}\).
The Sun and Earth were modeled as massive bodies, and the target and projectiles as massless test particles.
The gravitational parameters and the target's initial orbital elements were identical to Case~0.
The Sun--Earth relative orbit was circular, and its initial mean anomaly and that of the target orbit were both zero.

The simulations included \(1200\) projectiles with different initial states and the same semimajor axis as the Case~0 projectile.
For each projectile, a time was drawn uniformly at random within one LK cycle of the Case~0 projectile.
The initial eccentricity, inclination, and argument of periapsis were calculated from \((k_p,h_p)\) at that time using the Case~0 projectile's Kozai constant.
Each projectile's initial longitude of the ascending node and mean anomaly were drawn independently and uniformly at random on \([0,2\pi)\).

The integration used the Bulirsch--Stoer (BS) integrator, with both the absolute and relative error tolerances set to \(10^{-8}\).
The initial and maximum timesteps were both set to \(0.05\,\mathrm d\).
Collision detection used REBOUND's \texttt{LINETREE} mode, with the search restricted to encounters between the target and projectiles.
This mode approximates the motion of each body within an integration step as rectilinear motion at constant velocity.
A collision was identified if the minimum distance between the target and a projectile during the step did not exceed the effective collision radius \(R_{\mathrm{eff}}=100\,\mathrm{km}\).
When a collision was detected, the detection time was recorded and the corresponding projectile was removed.
The integration was terminated when the elapsed time reached \(1.5\times10^6\,\mathrm{yr}\) or all projectiles had been removed through collisions with the target.

The elapsed time since the start of the simulation is \(\tau=t-t_0\).
The number of projectiles that have not collided with the target by this time is denoted by \(N_{\mathrm{surv}}(\tau)\).
The empirical survival fraction from the REBOUND simulations is
\begin{equation}\label{p5res:empirical_survival}
S_{\mathrm{REB}}(\tau)
=
\frac{N_{\mathrm{surv}}(\tau)}{1200}.
\end{equation}
Approximating the collision frequency of each surviving projectile by the constant \(\Gamma_{\mathrm{Case}\,0}\) gives the semi-analytical survival prediction
\begin{equation}\label{p5res:theoretical_survival}
S_{\mathrm{th}}(\tau)
=
\exp\!\left(-\Gamma_{\mathrm{Case}\,0}\tau\right).
\end{equation}

\begin{figure}[tbp]
\centering
\includegraphics[width=0.92\linewidth]{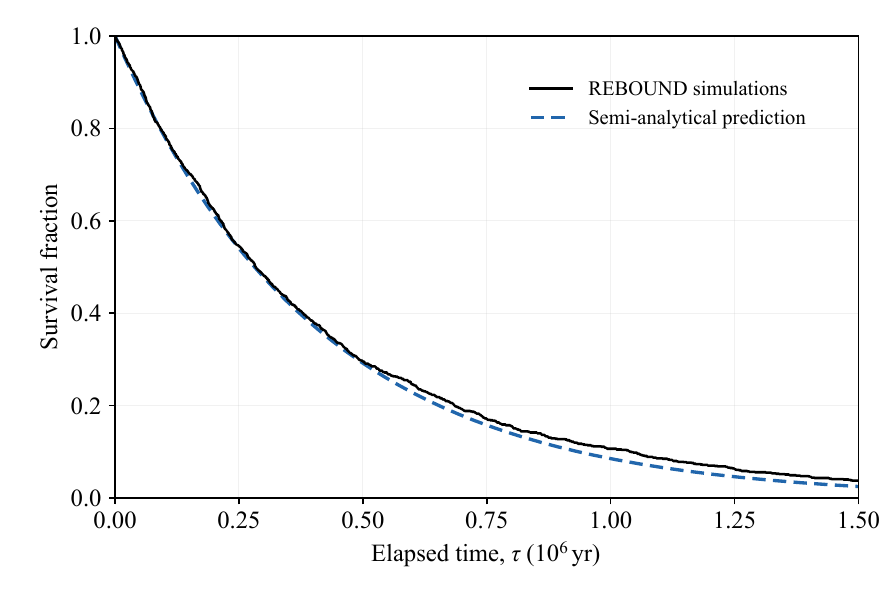}
\caption{
Comparison of REBOUND N-body simulation results with the semi-analytical prediction.
The blue dashed curve shows the semi-analytical prediction for Case~0, and the black solid curve shows the REBOUND simulation results.
}
\label{p5res:fig:survival}
\end{figure}

Figure~\ref{p5res:fig:survival} compares the projectile survival curve from the REBOUND simulations with the semi-analytical prediction.
The two curves show similar overall decay trends.
The fractions of projectiles removed by collisions at \(\tau=1.5\times10^6\,\mathrm{yr}\) were \(96.25\%\) in the REBOUND simulations and approximately \(97.51\%\) in the semi-analytical prediction.
These results indicate that, for the case considered, the framework provides a reasonable prediction of long-term projectile depletion through collisions.

\section{Conclusions}\label{p5:conclusion}

We have extended the existing \"Opik-type framework for calculating long-term average collision frequencies to cases in which both the projectile and target undergo LK evolution.
The secular evolution of both bodies is described by a doubly averaged quadrupole model, in which their semimajor axes remain constant and their orbital states evolve along their respective closed Hamiltonian level curves.
In the numerical example considered, treating the calculated collision frequency as constant yields a projectile survival curve with an overall decay trend close to that obtained from direct N-body simulations using REBOUND.

The theoretical analysis also indicates that, when \(1,Z_T,Z_\Omega\) are linearly independent over the rationals, changing the initial states of both bodies does not affect the long-term average collision frequency, provided that their semimajor axes, Kozai constants, and respective original closed Hamiltonian level curves remain fixed.
In the cases examined, varying both bodies' initial states under these constraints produced cumulative mean collision frequencies differing only slightly from the baseline estimate, consistent with the theoretical conclusion.

The framework could be used to estimate long-term average collision frequencies between irregular satellites orbiting giant planets and undergoing LK evolution.
The independence from initial states established here can help group large sets of numerical cases systematically, reducing repeated calculations.

\bibliographystyle{plainnat}
\bibliography{paper5_references}

\end{document}